# Achieving tunable surface tension in the pseudopotential lattice Boltzmann modeling of multiphase flows

Q. Li[1] and K. H. Luo[1, 2]

[1] Energy Technology Research Group, Faculty of Engineering and the Environment, University of Southampton, Southampton SO17 1BJ, United Kingdom

[2] Department of Mechanical Engineering, University College London, University of London, Torrington Place, London WC1E 7JE, United Kingdom

In this paper, we aim to address an important issue about the pseudopotential lattice Boltzmann (LB) model, which has attracted much attention as a mesoscopic model for simulating interfacial dynamics of complex fluids, but suffers from the problem that the surface tension cannot be tuned independently of the density ratio. In the literature, a multi-range potential was devised to adjust the surface tension [Sbragaglia *et al*., Phys. Rev. E **75**, 026702 (2007)]. However, it was recently found that the density ratio of the system will be changed when the multi-range potential is employed to adjust the surface tension. A new approach is therefore proposed in the present work. The basic strategy is to add a source term to the LB equation so as to tune the surface tension of the pseudopotential LB model. The proposed approach can guarantee that the adjustment of the surface tension does not affect the mechanical stability condition of the pseudopotential LB model, and thus provides a separate control of the surface tension and the density ratio. Meanwhile, it still retains the mesoscopic feature and the computational simplicity of the pseudopotential LB model. Numerical simulations are carried out for stationary droplets, capillary waves, and droplet splashing on a thin liquid film. The numerical results demonstrate that the proposed approach is capable of achieving a tunable surface tension over a very wide range and can keep the density ratio unchanged when adjusting the surface tension.

## I. Introduction

Interfaces between different phases or different components are ubiquitous in many complex fluids and soft matter systems. The dynamics of interfaces is difficult to study because the interfaces are usually deformable and their shapes are not known *a priori*. To track the motion of an interface, various numerical methods have been developed from different points of view. These methods are often classified into two categories [1], i.e., the interface-tracking approach, such as the front tracking method [2], and the interface-capturing approach, e.g., the volume of fluid method [3], the phase field method [4], and the level set method [5]. In the former approach, the interface is usually tracked with the motion of a cluster of marker points, while in the latter approach the interface is captured with the evolution of an order parameter, which is governed by an advection-diffusion equation.

In the past decade, the lattice Boltzmann (LB) method, which is a mesoscopic approach based on minimal lattice formulations of the kinetic Boltzmann equation [6-9], has attracted considerable attention as an efficient alternative to traditional methods for modeling interfacial phenomena in multiphase and multicomponent flows [10-19]. A very popular multiphase model is the pseudopotential LB model proposed by Shan and Chen [10]. In this model, the intermolecular interactions are represented with a density-dependent pseudopotential, and the phase separation is achieved by imposing a short-range attraction between different phases. Consequently, interfaces can arise, deform, and migrate naturally in the pseudopotential LB modeling without using a cluster of marker points to track the interfaces or capturing the interfaces via the evolution of an order parameter [17].

Because of its mesoscopic feature and remarkable computational simplicity, the pseudopotential LB model has captured significant interest since its emergence, and has been successfully applied in modeling multiphase/multicomponent flows and soft-flowing materials [20, 21]. However, the basic model has some drawbacks, such as large spurious currents in the presence of large density ratio. This problem has been extensively studied in the literature, and it has been found that the spurious currents can be reduced by using high-order isotropic gradient operators [22] or enlarging the interface thickness (in terms of lattice units) [23]. Another drawback of the pseudopotential LB model is that the surface tension, which plays an important role in interfacial phenomena, cannot be tuned independently of the density ratio. In order to make the surface tension adjustable, Sbragaglia *et al.* [23] have developed a multi-range potential through combining the nearest-neighbor interactions and the next-nearest-neighbor interactions, which leads to a two-parameter equation of state, and therefore allows a separate control of the equation of state and the surface tension. This approach has also been applied to higher-order lattices to construct a two-belt pseudopotential LB model [24, 25]. In Ref. [23], it was pointed out that the multi-range potential can push the pseudopotential LB model at varying the density ratios and the surface tensions independently [23]. Nevertheless, Huang *et al.* recently found that [26], using the multi-range potential, the density ratio of the system will considerably vary with the surface tension.

In the present work, through theoretical and numerical analyses of the multi-range potential, we will elucidate the reason why the density ratio changes when the multi-range potential is used to tune the surface tension. Subsequently, an alternative approach will be developed to tune the surface tension of the pseudopotential LB model. The proposed approach is capable of adjusting the surface tension independently of the density ratio. The rest of the present paper is organized as follows. Section II will briefly introduce the pseudopotential LB model. Theoretical and numerical analyses of the multi-range potential will be conducted in Section III. The new approach to adjusting the surface tension of the pseudopotential LB model will be presented in Section IV. Numerical results will be shown in Section V. Finally, a brief conclusion will be made in Section VI.

## II. The pseudopotential LB model

Without loss of generality, we consider a two-dimensional nine-velocity (D2Q9) pseudopotential LB model with a multiple-relaxation-time (MRT) collision operator [27, 28]. The MRT LB equation can be written as

$$f_\alpha \left( \mathbf{x} + \mathbf{e}_\alpha \delta_t, t + \delta_t \right) = f_\alpha \left( \mathbf{x}, t \right) - \left( \mathbf{M}^{-1} \mathbf{\Lambda} \mathbf{M} \right)_{\alpha\beta} \left( f_\beta - f_\beta^{eq} \right) + \delta_t F'_\alpha, \qquad (1)$$

where $f_\alpha$ is the discrete single-particle density distribution function, $\mathbf{e}_\alpha = \left( e_{\alpha x}, e_{\alpha y} \right)$ is the discrete velocity in the $\alpha$-direction, $\delta_t$ is the time step, $F'_\alpha$ represents the forcing term in the velocity space, $\mathbf{\Lambda} = \mathrm{diag}\left( \tau_\rho^{-1}, \tau_e^{-1}, \tau_\varsigma^{-1}, \tau_j^{-1}, \tau_q^{-1}, \tau_j^{-1}, \tau_q^{-1}, \tau_v^{-1}, \tau_v^{-1} \right)$ is the diagonal Matrix, and $\mathbf{M}$ is the orthogonal transformation matrix. Using the transformation matrix, the right hand side of Eq. (1) can be rewritten as [29]

$$\mathbf{m}^* = \mathbf{m} - \mathbf{\Lambda}\left( \mathbf{m} - \mathbf{m}^{eq} \right) + \delta_t \left( \mathbf{I} - \frac{\mathbf{\Lambda}}{2} \right) \mathbf{S}, \qquad (2)$$

where $\mathbf{I}$ is the unit tensor, $\mathbf{S}$ is the forcing term in the moment space with $(\mathbf{I} - 0.5\mathbf{\Lambda})\mathbf{S} = \mathbf{MF}'$, and the equilibria $\mathbf{m}^{eq}$ is given by

$$\mathbf{m}^{eq} = \rho\left( 1, -2 + 3|\mathbf{v}|^2, 1 - 3|\mathbf{v}|^2, v_x, -v_x, v_y, -v_y, v_x^2 - v_y^2, v_x v_y \right)^{\mathrm{T}}. \qquad (3)$$

The streaming process of the MRT LB equation is given as

$$f_\alpha \left( \mathbf{x} + \mathbf{e}_\alpha \delta_t, t + \delta_t \right) = f_\alpha^* \left( \mathbf{x}, t \right), \qquad (4)$$

where $\mathbf{f}^* = \mathbf{M}^{-1} \mathbf{m}^*$. The forcing terms in the moment space, $\mathbf{S}$, are given by

$$S_0 = 0, \quad S_1 = 6\left( v_x F_x + v_y F_y \right), \quad S_2 = -S_1, \quad S_3 = F_x, \quad S_4 = -F_x,$$

$$S_5 = F_y, \quad S_6 = -F_y, \quad S_7 = 2(v_x F_x - v_y F_y), \quad S_8 = (v_x F_y + v_y F_x). \tag{5}$$

The corresponding macroscopic density and velocity are calculated via

$$\rho = \sum_\alpha f_\alpha, \quad \rho \mathbf{v} = \sum_\alpha \mathbf{e}_\alpha f_\alpha + \frac{\delta_t}{2}\mathbf{F}, \tag{6}$$

where $\mathbf{F} = (F_x, F_y)$ is the interaction force [30]

$$\mathbf{F} = -G\psi(\mathbf{x})\sum_{\alpha=1}^{8} w(|\mathbf{e}_\alpha|^2)\psi(\mathbf{x}+\mathbf{e}_\alpha)\mathbf{e}_\alpha, \tag{7}$$

in which $\psi$ is the interaction potential, $G$ is the interaction strength, and $w(|\mathbf{e}_\alpha|^2)$ are the weights. For the nearest-neighbor interactions on the D2Q9 lattice, the weights are $w(1) = 1/3$ and $w(2) = 1/12$.

In the original pseudopotential LB model proposed by Shan and Chen [10], the interaction potential $\psi$ is chosen as $\psi(\rho) = \psi_0 \exp(-\rho_0/\rho)$. For this choice, the thermodynamic consistency can be guaranteed with the forcing scheme Eq. (5). However, such an interaction potential is usually limited to low-density-ratio interfacial problems and does not allow a prescribed equation of state. Another potential, $\psi = \sqrt{2(p_{\mathrm{EOS}} - \rho c_s^2)/Gc^2}$, was therefore proposed in the literature [14, 31]. Here $p_{\mathrm{EOS}}$ represents a prescribed equation of state (e.g., the Carnahan-Starling equation of state). Nevertheless, using this type of interaction potential, the pseudopotential LB model will suffer from the lack of thermodynamic consistency, namely the coexistence densities given by the pseudopotential LB model will be inconsistent with those given by the *Maxwell construction* [7]. Recently, we found [32] that in this case the thermodynamic consistency can be approximately achieved through adjusting the mechanical stability condition via an improved forcing scheme (BGK form in [32] and MRT form in [33]). The improved MRT forcing terms are the same as those in Eq. (5) except that $S_1$ is modified as [33]

$$S_1 = 6(v_x F_x + v_y F_y) + \frac{12\sigma|\mathbf{F}|^2}{\psi^2 \delta_t (\tau_e - 0.5)}, \tag{8}$$

where $|\mathbf{F}|^2 = (F_x^2 + F_y^2)$ and $\sigma$ is a constant. The forcing term $S_2$ is still defined as $S_2 = -S_1$ with the non-dimensional relaxation time $\tau_\varsigma = \tau_e$.

## III. Analyses of the multi-range potential

### A. Theoretical analysis

In this section, theoretical and numerical analyses will be carried out for the multi-range potential proposed by Sbragaglia *et al*. [23], which is given by

$$\mathbf{F} = -\psi(\mathbf{x}) \sum_{\alpha=1}^{N} w(|\mathbf{e}_\alpha|^2) \left[ G_1 \psi(\mathbf{x}+\mathbf{e}_\alpha) + G_2 \psi(\mathbf{x}+2\mathbf{e}_\alpha) \right] \mathbf{e}_\alpha . \tag{9}$$

Using the Taylor expansion, Sbragaglia *et al*. have obtained the following pressure tensor for the multi-range potential:

$$\mathbf{P}_c = \left[ p_0 + (G_1 + 8G_2) \left( \frac{c^4}{12} |\nabla \psi|^2 + \frac{c^4}{6} \psi \nabla^2 \psi \right) \right] \mathbf{I} - \frac{(G_1 + 8G_2)c^4}{6} \nabla \psi \nabla \psi , \tag{10}$$

where $p_0(\rho) = \rho c_s^2 + (G_1 + 2G_2) c^2 \psi^2 / 2$ is the equation of state. According to Eq. (10), the surface tension obtained with the multi-range potential should be close to zero when $(G_1 + 8G_2) \sim 0$.

Equation (10) is the *continuum* form pressure tensor of the multi-range potential. However, Shan [30] has argued that, for the pseudopotential LB model, the *discrete* form pressure tensor must be used in order to guarantee the exact mechanical balance. Here we derive the discrete form pressure tensor of the multi-range potential following the line of Ref. [30]. By noticing that $G_2 \psi(\mathbf{x}+2\mathbf{e}_\alpha)\mathbf{e}_\alpha$ in Eq. (9) can be rewritten as $0.5 G_2 \psi(\mathbf{x}+2\mathbf{e}_\alpha)(2\mathbf{e}_\alpha)$, in which $2\mathbf{e}_\alpha$ is equivalent to $\mathbf{e}_{\alpha=9,10,\cdots,16}$ in Ref. [30], we can evaluate the discrete form pressure tensor of the multi-range potential as follows:

$$\mathbf{P} = \rho c_s^2 \mathbf{I} + \frac{G_1}{2} \psi(\mathbf{x}) \sum_\alpha w(|\mathbf{e}_\alpha|^2) \psi(\mathbf{x}+\mathbf{e}_\alpha) \mathbf{e}_\alpha \mathbf{e}_\alpha + \frac{0.5 G_2}{4} \psi(\mathbf{x}) \sum_\alpha w(|\mathbf{e}_\alpha|^2) \psi(\mathbf{x}+2\mathbf{e}_\alpha)(2\mathbf{e}_\alpha)(2\mathbf{e}_\alpha)$$

$$+ \frac{0.5 G_2}{4} \sum_\alpha w(|\mathbf{e}_\alpha|^2) \psi\left(\mathbf{x}+\frac{2\mathbf{e}_\alpha}{2}\right) \psi\left(\mathbf{x}-\frac{2\mathbf{e}_\alpha}{2}\right)(2\mathbf{e}_\alpha)(2\mathbf{e}_\alpha) . \tag{11}$$

When $G_1 = G$ and $G_2 = 0$, Eq. (11) gives the standard pressure tensor of the pseudopotential LB model:

$$\mathbf{P} = \left( \rho c_s^2 + \frac{Gc^2}{2} \psi^2 + \frac{Gc^4}{12} \psi \nabla^2 \psi \right) \mathbf{I} + \frac{Gc^4}{6} \psi \nabla \nabla \psi . \tag{12}$$

For general cases, the following equation can be obtained from Eq. (11) after some standard algebra:

$$\mathbf{P} = \left( p_0 + \frac{(G_1 + 6G_2)c^4}{12} \psi \nabla^2 \psi - \frac{G_2 c^4}{6} |\nabla \psi|^2 \right) \mathbf{I} + \frac{(G_1 + 6G_2)c^4}{6} \psi \nabla \nabla \psi - \frac{G_2 c^4}{3} \nabla \psi \nabla \psi. \quad (13)$$

Here the equation of state is also given by $p_0(\rho) = \rho c_s^2 + (G_1 + 2G_2)c^2 \psi^2 / 2$. Note that higher-order terms have been neglected in Eq. (13).

From Eq. (13) it can be seen that the discrete form pressure tensor of the multi-range potential contains two terms related to the surface tension: the last two terms on the right-hand side of Eq. (13). These two terms can be rewritten as follows:

$$\frac{(G_1 + 6G_2)c^4}{6} \psi \nabla \nabla \psi - \frac{G_2 c^4}{3} \nabla \psi \nabla \psi = \frac{c^4}{6} \left[ (G_1 + 8G_2) \psi \nabla \nabla \psi - 2G_2 \nabla (\psi \nabla \psi) \right]. \quad (14)$$

It is obvious that, as long as $G_2$ is non-negligible, the surface tension given by the multi-range potential will not approach zero when $(G_1 + 8G_2) \sim 0$.

Furthermore, it can be proven that Eq. (13) will lead to the following normal pressure tensor for a flat interface:

$$P_n = p_0 + \frac{(G_1 + 6G_2)c^4}{12} \left[ -\frac{6G_2}{(G_1 + 6G_2)} \left( \frac{d\psi}{dn} \right)^2 + 3\psi \frac{d^2\psi}{dn^2} \right], \quad (15)$$

where $n$ denotes the normal direction of the interface. According to Eq. (15) and the requirement that at equilibrium $P_n$ should be equal to the constant static pressure in the bulk [30], the mechanical stability condition will be given by

$$\int_{\rho_V}^{\rho_L} \left( p_b - \rho c_s^2 - \frac{(G_1 + 2G_2)c^2}{2} \psi^2 \right) \frac{\psi'}{\psi^{1+\varepsilon}} d\rho = 0, \quad \varepsilon = \frac{4G_2}{G_1 + 6G_2}, \quad (16)$$

where $\rho_L$ and $\rho_V$ are the coexistence liquid and vapor densities, and $p_b$ is the bulk pressure.

For the pseudopotential LB model, the coexistence densities are not only related to the equation of state, but also affected by the mechanical stability condition. From Eq. (16) it can be seen that the mechanical stability condition given by the multi-range potential is dependant on $\varepsilon = 4G_2 / (G_1 + 6G_2)$, which means that the coexistence densities obtained by the multi-range potential will vary with $G_1$ and $G_2$. This is the reason why the density ratio of the system will be changed when the multi-range potential is utilized to adjust the surface tension. Nevertheless, it should also be recognized that the multi-range potential has been applied with considerable success in many applications, e.g., the simulation of disjoining pressure [25], and was found to be a promising tool for modeling complex flow phenomena, such as spray formation, micro-emulsions, and breakup phenomena [43].

### B. Numerical analysis

Now numerical analysis is performed for the multi-range potential. A stationary droplet is considered. The computational domain is taken as $N_x \times N_y = 120 \times 120$ and a circular droplet with a radius of $R = 40$ is initially placed at the center of the domain with the liquid phase inside the droplet. The periodical boundary conditions are applied in the $x$- and $y$-directions. The kinematic viscosity is set to be $\upsilon = 0.1$. In order to test whether the surface tension will approach zero when $(G_1 + 8G_2) \sim 0$, the following four cases are considered:

(A) $G_1 = G$ and $G_2 = 0$; (B) $G_1 = 1.15G$ and $G_2 = -0.075G$;

(C) $G_1 = 1.3G$ and $G_2 = -0.15G$; (D) $G_1 = 1.33G$ and $G_2 = -0.165G$.

With these choices, $(G_1 + 2G_2)$ is fixed at $G$. Then the equation of state is given by $p(\rho) = \rho c_s^2 + Gc^2\psi^2/2$ for the four cases. The corresponding parameter $(G_1 + 8G_2)$ is equal to $G$, $0.55G$, $0.1G$, and $0.01G$, respectively.

The interaction potential is chosen as $\psi = \psi_0 \exp(-\rho_0/\rho)$ with $\psi_0 = 4$ and $\rho_0 = 200$ [7]. The interaction strength $G$ in Eq. (7) is set to be $G = -40$. For such a choice, the liquid and vapor densities given by the Maxwell construction are $\rho_L \approx 514$ and $\rho_V \approx 79.5$, respectively. The numerical coexistence densities of the four cases are plotted in Fig. 1, from which it can be seen that both the liquid and vapor densities considerably change with the parameter $(G_1 + 8G_2)$. In particular, at Case D the relative errors of the vapor density and the density ratio are about 10% and 9%, respectively, which confirms that the multi-range potential cannot keep the coexistence densities and the density ratio unchanged.

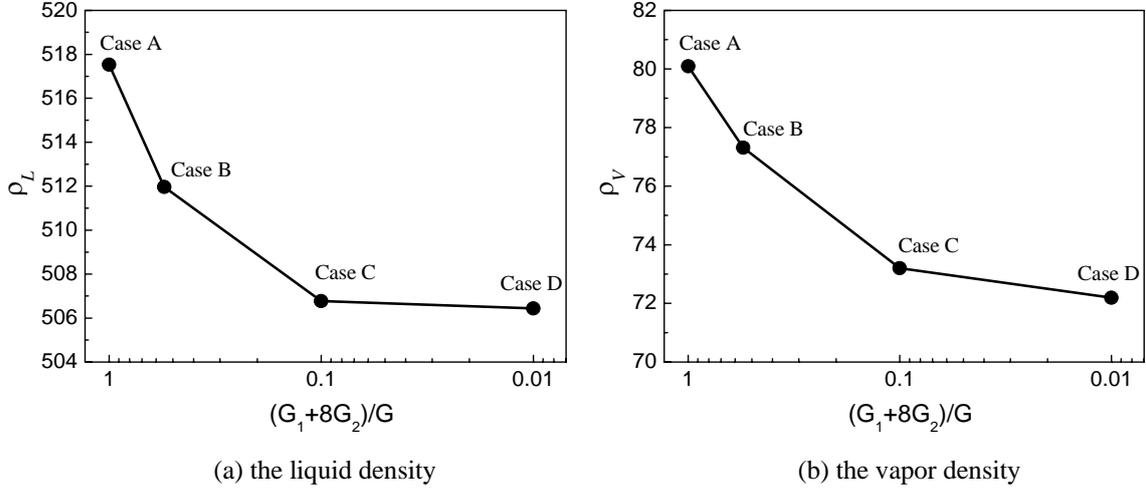

(a) the liquid density  (b) the vapor density

**Figure 1** The coexistence densities obtained with the multi-range potential.

The surface tensions (evaluated with the Laplace's law) of the four cases are displayed in Table 1. As can be seen in the table, the surface tension given by the multi-range potential does not approach zero when $(G_1 + 8G_2) \sim 0$. From Case A to Case D, the parameter $(G_1 + 8G_2)$ changes from $G$ to $0.01G$, however, the surface tension is only lowered to 3.971 from 9.449. Furthermore, it can be seen that the results of Cases C and D are just slightly different, although the parameter $(G_1 + 8G_2)$ of Case C is ten times larger than that of Case D. With the numerical results, it can be concluded that the multi-range potential does not obey the continuum pressure tensor given by Eq. (10). On the other hand, we find that the numerical results can be well illustrated with the discrete form pressure tensor. According to Eq. (14), the detailed forms of $\left[\left(G_1+8G_2\right)\psi\nabla\nabla\psi-2G_2\nabla\left(\psi\nabla\psi\right)\right]$ are also listed in Table 1. From Case A to Case D, with the decrease of $\left(G_1+8G_2\right)$, the surface tension is reduced via the term $\left(G_1+8G_2\right)\psi\nabla\nabla\psi$. However, for Cases B, C, and D, another term $\left[-2G_2\nabla\left(\psi\nabla\psi\right)\right]$ appears and gradually enhances the surface tension. As a result, the surface tension is limited in a narrow range. This is the reason why the surface tensions of the four cases are of the same order of magnitude.

**Table 1** The surface tension obtained with the multi-range potential.

| Case | $G_1 + 8G_2$ | surface tension | $(G_1 + 8G_2)\psi\nabla\nabla\psi - 2G_2\nabla(\psi\nabla\psi)$ |
|------|--------------|-----------------|------------------------------------------------------------------|
| A    | $G$          | 9.449           | $G\psi\nabla\nabla\psi$                                          |
| B    | $0.55G$      | 7.181           | $0.55G\psi\nabla\nabla\psi + 0.15G\nabla(\psi\nabla\psi)$        |
| C    | $0.1G$       | 4.639           | $0.1G\psi\nabla\nabla\psi + 0.3G\nabla(\psi\nabla\psi)$          |
| D    | $0.01G$      | 3.971           | $0.01G\psi\nabla\nabla\psi + 0.33G\nabla(\psi\nabla\psi)$        |

Actually, the surface tension of Case D can be estimated from the results of the former three cases by assuming that the surface tensions resulting from $G\psi\nabla\nabla\psi$ and $G\nabla\left(\psi\nabla\psi\right)$ do not change among the considered cases. According to such an assumption and the numerical results of Cases B and C, it will be found that $G\nabla\left(\psi\nabla\psi\right)$ represents a surface tension around 12.222. Then the surface tension of Case D can be approximately estimated as: $\vartheta_D = 0.01\times 9.449 + 0.33\times 12.222 \approx 4.128$. It can be seen that the estimated result is very close to the corresponding numerical result in Table 1.

## IV. New approach to adjusting the surface tension

In the previous section, we have conducted theoretical and numerical analyses for the multi-range potential. The discrete form pressure tensor of the multi-range potential has been derived and it has been found that, when the multi-range potential is used to tune the surface tension, the mechanical stability condition of the pseudopotential LB model will vary with the parameters $G_1$ and $G_2$, which will result in considerable changes of the coexistence densities and the density ratio.

In order to overcome the weakness of the multi-range potential, we propose an alternative approach to adjust the surface tension of the pseudopotential LB model, which is still based on the *single-range* potential and is implemented by adding a source term to the MRT LB equation:

$$\mathbf{m}^* = \mathbf{m} - \Lambda(\mathbf{m} - \mathbf{m}^{eq}) + \delta_t\left(\mathbf{I} - \frac{\Lambda}{2}\right)\mathbf{S} + \delta_t\mathbf{C}, \tag{17}$$

where the source term $\mathbf{C}$ is given by

$$\mathbf{C} = \begin{bmatrix} 0 \\ 1.5\tau_e^{-1}(Q_{xx} + Q_{yy}) \\ -1.5\tau_\varsigma^{-1}(Q_{xx} + Q_{yy}) \\ 0 \\ 0 \\ 0 \\ 0 \\ -\tau_v^{-1}(Q_{xx} - Q_{yy}) \\ -\tau_v^{-1}Q_{xy} \end{bmatrix}. \tag{18}$$

Note that the discrete effect [34] of the source term has been considered and the related terms have been incorporated into the source term. The variables $Q_{xx}$, $Q_{yy}$, and $Q_{xy}$ are calculated via

$$\mathbf{Q} = \kappa\frac{G}{2}\psi(\mathbf{x})\sum_{\alpha=1}^{8}w(|\mathbf{e}_\alpha|^2)[\psi(\mathbf{x}+\mathbf{e}_\alpha) - \psi(\mathbf{x})]\mathbf{e}_\alpha\mathbf{e}_\alpha, \tag{19}$$

where the parameter $\kappa$ is used to tune the surface tension. In fact, $\mathbf{Q}$ is based on the discrete form pressure tensor [30] of the standard pseudopotential LB model, which can be formulated as

$$\begin{aligned}\mathbf{P} &= \rho c_s^2\mathbf{I} + \frac{G}{2}\psi(\mathbf{x})\sum_{\alpha=1}^{8}w(|\mathbf{e}_\alpha|^2)\psi(\mathbf{x}+\mathbf{e}_\alpha)\mathbf{e}_\alpha\mathbf{e}_\alpha \\ &= \left(\rho c_s^2 + \frac{Gc^2}{2}\psi^2\right)\mathbf{I} + \frac{G}{2}\psi(\mathbf{x})\sum_{\alpha=1}^{8}w(|\mathbf{e}_\alpha|^2)[\psi(\mathbf{x}+\mathbf{e}_\alpha) - \psi(\mathbf{x})]\mathbf{e}_\alpha\mathbf{e}_\alpha.\end{aligned} \tag{20}$$

Through the Chapman-Enskog analysis [35, 28], one can find that an additional term will be introduced into the Navier-Stokes (N-S) equations:

$$\text{N-S}_{\text{new}} = \text{N-S}_{\text{original}} - \nabla\cdot\left[\kappa\frac{Gc^4}{6}(\psi\nabla^2\psi\mathbf{I} - \psi\nabla\nabla\psi)\right]. \tag{21}$$

Accordingly, the discrete form pressure tensor will be given by

$$\mathbf{P}_{\text{new}} = \left[\rho c_s^2 + \frac{Gc^2}{2}\psi^2 + \frac{Gc^4}{12}(1+2\kappa)\psi\nabla^2\psi\right]\mathbf{I} + \frac{Gc^4}{6}(1-\kappa)\psi\nabla\nabla\psi. \tag{22}$$

When $\kappa = 0$, Eq. (22) will reduce to the standard pressure tensor of the pseudopotential LB model.

Now some statements are made about the proposed new approach. First, as can be seen in Eq. (22), the coefficient in front of the term related to the surface tension has been changed from 1 to $(1-\kappa)$. The surface tension is therefore expected to decrease and approach zero when the parameter $\kappa$ increases from 0 to 1. Moreover, from Eq. (22) it can be found that an additional term $\kappa G c^4 \psi \nabla^2 \psi \mathbf{I}/6$ has also been introduced into the new pressure tensor. This term is utilized to ensure that the adjustment of the surface tension will not affect the mechanical stability condition of the pseudopotential LB model, and thus guarantees that the surface tension can be tuned independently of the density ratio.

Finally, we would like to point out that, although the adjustment of the surface tension is pretty easy in some other LB models, it doesn't mean the pseudopotential LB model is inferior to these models. The remarkable advantages of the pseudopotential LB model have been stated previously. By comparing Eq. (19) with Eq. (7), we can see that the mesoscopic feature and the computational simplicity of the pseudopotential LB model are retained in the proposed new approach.

## V. Numerical results

In this section, numerical simulations will be carried out for stationary droplets, capillary waves, and droplet splashing on a thin liquid film to validate the proposed approach.

### A. Stationary droplets

To start with, the problem of stationary droplets is considered. For the sake of enabling a comparison between the present results and the results in Section 3.2, we first adopt the interaction potential used in Section 3.2, i.e., $\psi = \psi_0 \exp(-\rho_0/\rho)$. The settings, including the lattice size, the initial droplet radius, the relaxation times, etc., are the same as those in Section 3.2. The parameter $\kappa$ in Eq. (19) increases from 0 to 0.99. Correspondingly, $(1-\kappa)$ decreases from 1 to 0.01.

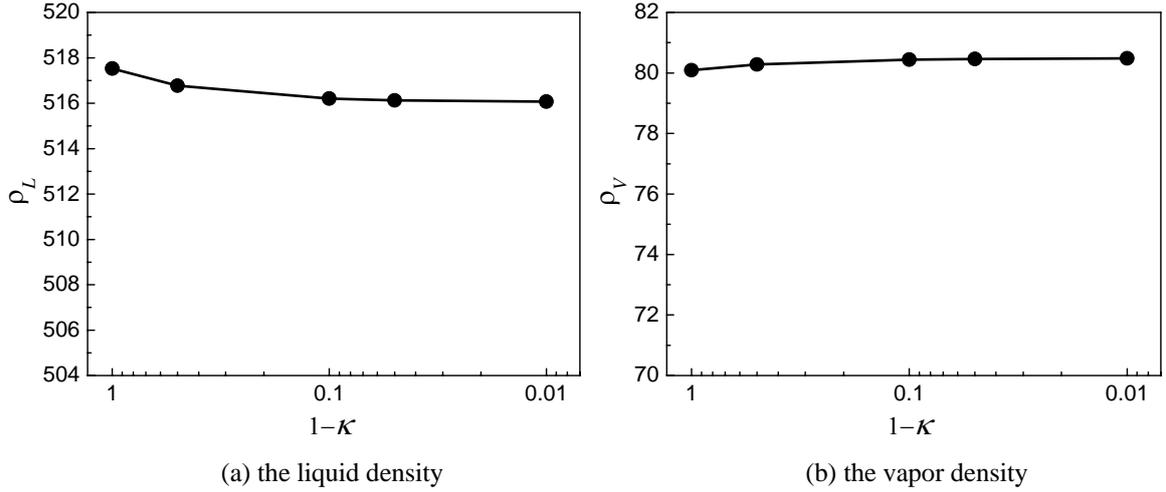

(a) the liquid density  (b) the vapor density

**Figure 2** The coexistence densities obtained with the proposed new approach ($\psi = \psi_0 e^{-\rho_0/\rho}$).

The numerical coexistence densities are displayed in Fig. 2 as a function of $(1-\kappa)$. By comparing Fig. 2 with Fig. 1, we can see that the coexistence densities given by the new approach remain basically unchanged. To be specific, when $(1-\kappa)$ decreases from 1 to 0.01, the liquid (vapor) density changes from 517.5 (80.1) to 516.1 (80.5), with the density ratio varying from 6.461 to 6.411. The variations are mainly attributed to the neglected higher-order terms in the pressure tensor, and it can be seen that the relative variation of the density ratio is 0.8%, which is negligible. The numerical surface tensions at $\kappa = 0$, 0.5, 0.95, and 0.99 are listed in Table 2. It is observed that, when $(1-\kappa)$ varies from 1 to 0.01, the surface tension gradually approaches zero. Particularly, the surface tension of the case $\kappa = 0.99$ is about $1/266$ of the surface tension at $\kappa = 0$ and is far smaller than that given by the multi-range potential with $(G_1 + 8G_2) = 0.01G$. To sum up, the numerical results demonstrate that the proposed approach is capable of adjusting the surface tension over a wide range with an unchanged density ratio.

**Table 2** The surface tension obtained with the proposed new approach ($\psi = \psi_0 e^{-\rho_0/\rho}$).

| $\kappa$ | 0 | 0.5 | 0.95 | 0.99 |
|---|---|---|---|---|
| $1-\kappa$ | 1 | 0.5 | 0.05 | 0.01 |
| surface tension | 9.449 | 4.605 | 0.394 | 0.0355 |

Furthermore, the interaction potential $\psi = \sqrt{2(p_{EOS} - \rho c_s^2)/Gc^2}$ is also considered. A piecewise linear equation of state, which is recently developed by Colosqui *et al.* [36], is adopted, i.e.,

$$p_{EOS} = \begin{cases} \rho \theta_V & \text{if } \rho \leq \rho_1 \\ \rho_1 \theta_V + (\rho - \rho_1)\theta_M & \text{if } \rho_1 < \rho < \rho_2, \\ \rho_1 \theta_V + (\rho_2 - \rho_1)\theta_M + (\rho - \rho_2)\theta_L & \text{if } \rho \geq \rho_2 \end{cases} \tag{23}$$

where $\sqrt{\theta_V} = \sqrt{(\partial p/\partial \rho)_V}$ and $\sqrt{\theta_L} = \sqrt{(\partial p/\partial \rho)_L}$ are the vapor-phase and liquid-phase speed of sound, respectively, and $\theta_M$ is the slope in the unstable branch ($\partial p/\partial \rho < 0$). Compared with the classical equations of state, such as the van der Waals and the Carnahan-Starling equations of state, the equation of state defined by Eq. (23) can offer a control of $\partial p/\partial \rho$ in different regions. In addition, a desired density ratio can be readily prescribed by such an equation of state.

In simulations, the density ratio is set to be 100 with $\rho_L = 100$ and $\rho_V = 1$, and the parameters are chosen as $\theta_V = 0.49c_s^2$, $\theta_L = c_s^2$, and $\theta_M = -0.06c_s^2$, which gives an interface width about five lattices. Adopting an interface thickness around 4-5 lattices can reduces the spurious currents of the pseudopotential LB model as compared with an interface thickness around 2 ~ 3 lattices. The parameter $G$ is set to be $-1.0$. According to the mechanical equilibrium and the chemical equilibrium (see Ref. [36] for details), the variables $\rho_1$ and $\rho_2$ in Eq. (23) are given by $\rho_1 = 1.49$ and $\rho_2 = 94.65$, respectively. Previously we have mentioned that, when the potential $\psi = \sqrt{2(p_{EOS} - \rho c_s^2)/Gc^2}$ is used, the pseudopotential LB model will suffer from the problem of thermodynamic inconsistency. To eliminate the thermodynamic inconsistency, the improved forcing scheme Eq. (8) is employed. The parameter $\sigma$ in Eq. (8) is set to be 0.087.

Table 3 The numerical results obtained with the proposed new approach.

| $1-\kappa$ | $\rho_V$ | $\rho_L$ | surface tension |
|---|---|---|---|
| 1 | 1.005 | 100.14 | 1.6401 |
| 0.5 | 1.012 | 100.08 | 0.8160 |
| 0.05 | 1.019 | 100.06 | 0.0830 |
| 0.01 | 1.019 | 100.06 | 0.0215 |

The numerical coexistence densities and surface tensions at $1-\kappa = 1$, $0.5$, $0.05$, and $0.01$ ($R = 40$) are shown in Table 3. From the table we can see that there are only minor variations for the liquid and vapor densities when $1-\kappa$ changes from 1 to 0.01. Meanwhile, it is seen that the predicted surface tension gradually approaches zero when $1-\kappa \sim 0$. To validate the Laplace's law, droplets with different radii are simulated. According to the Laplace's law, the pressure difference across a circular interface is related to the surface tension $\vartheta$ and the droplet radius $R$ via $\delta p = p_{in} - p_{out} = \vartheta/R$. When the surface tension is given, the pressure difference $\delta p$ will be proportional to $1/R$. The numerical pressure differences at $1-\kappa = 1$, $0.5$, and $0.05$ with $20 \leq R \leq 40$ are shown in Fig. 3. It can be seen that the numerical results agree well with the linear fit denoted by the solid lines.

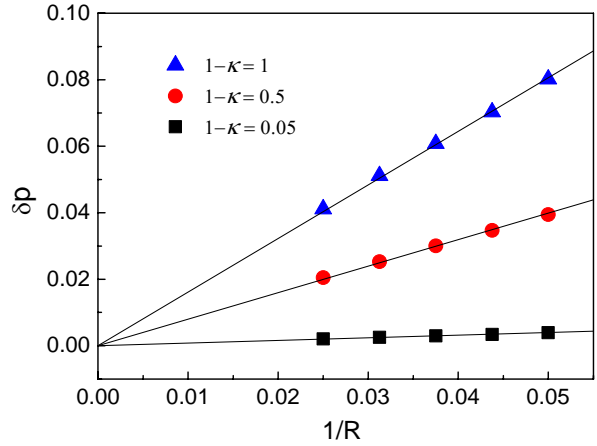

**Figure 3** Numerical validation of the Laplace's law.

## B. Capillary waves

To verify the dynamic behavior of the proposed approach, numerical simulations are carried out for capillary waves between two fluids with equal viscosities. The damped oscillation of capillary waves is a classical test of the accuracy of numerical schemes for simulating surface-tension-driven interfacial dynamics. The simulations are carried out on a rectangular domain with length $L$ and height $H$. To reduce the effects of finite water depth [10], the aspect ratio of the domain $H/L$ is chosen as $3.5$ with $L = 160$. The periodic boundary condition is adopted in the $x$-direction and the non-slip boundary condition is applied in the $y$-direction.

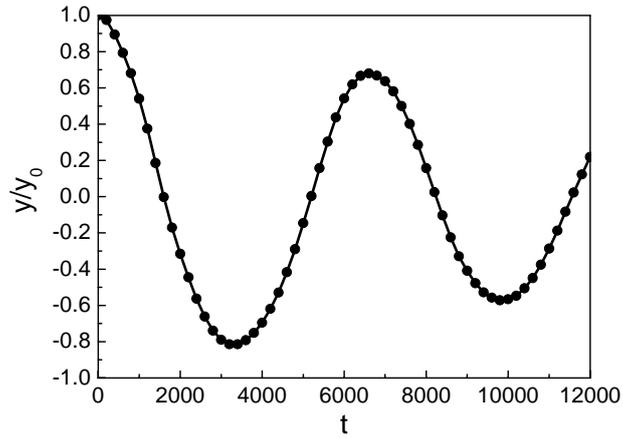

(a) $1-\kappa = 1$

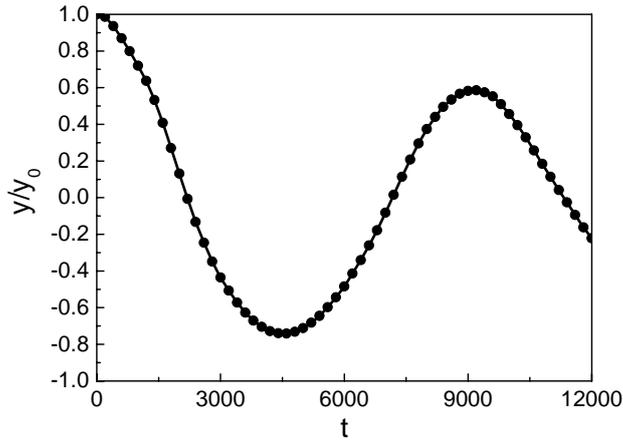

(b) $1-\kappa = 0.5$

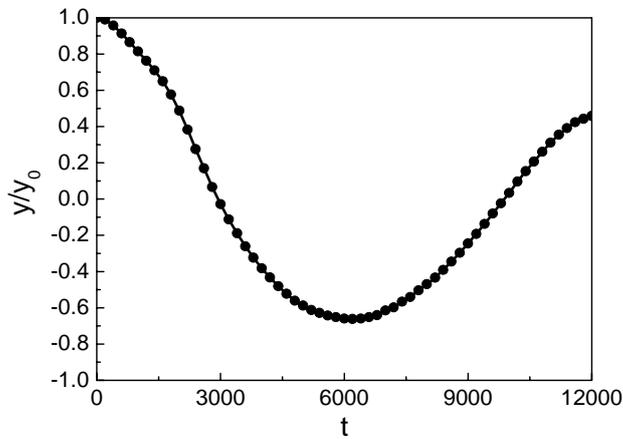

(c) $1-\kappa = 0.25$

**Figure 4** The evolution of the capillary wave amplitude at $1-\kappa = 1$, $0.5$, and $0.25$.

A sinusoidal wave is initially placed in the middle of the domain $[0, L] \times [-H/2, H/2]$ and located at $y(x) = h_0 \cos(kx)$, where $k = 2\pi/L$ is the wave number and $h_0$ is the wave amplitude, which should be much larger than the interface thickness and is set to be $0.125L$. When the kinematic viscosities of the two fluids are identical, the dispersion relation for capillary waves is given by

$$\omega = \frac{\vartheta k^3}{\rho_L + \rho_V}, \tag{24}$$

where $\omega$ is the oscillating frequency. The oscillating period is defined as $T = 2\pi/\omega$. In simulations, the interaction potential $\psi = \sqrt{2(p_{EOS} - \rho c_s^2)/Gc^2}$ together with Eq. (23) is adopted. The densities are chosen as $\rho_L = 100$ and $\rho_V = 1$. The kinematic viscosities $\upsilon_L$ and $\upsilon_V$ are both set to be 0.01. Three different cases are considered: $1-\kappa = 1$, $0.5$, and $0.25$. Figure 4 shows the evolution of the capillary wave amplitude between $t = 0$ and $t = 12000\delta_t$ ($\delta_t = 1$). The results are measured in the interface $\rho = (\rho_L + \rho_V)/2$ and $y_0$ is the wave amplitude at $t = 0$. From the figure we can see that the dynamic behavior of capillary waves is well described by the proposed approach. In addition, it can be seen that the oscillating period increases when the surface tension decreases. To quantify the results, the numerical oscillating periods are compared with the corresponding analytical results in Table 4. As can be seen in the table, the present results are basically in good agreement with the analytical solution. The minor discrepancies may be attributed to the fact that the analytical solution is derived for sharp interfaces.

**Table 4** Comparison of the oscillating period between the numerical and analytical results.

| $1-\kappa$ | 1 | 0.5 | 0.25 |
|---|---|---|---|
| analytical | 6389.6 | 9072.1 | 12793.4 |
| numerical | 6600 | 9150 | 12350 |
| error (%) | 3.3 | 0.9 | 3.5 |

### C. Droplet splashing on a thin liquid film

The splashing phenomenon that occurs after liquid droplets impact onto a solid or liquid surface can be observed in everyday scenarios, such as raindrops falling on the ground and pouring the morning coffee, and can be found in a wide variety of technological applications, e.g., spray coating and cooling [37], printing [38], droplet impacting on superheated surfaces [39], and the impact of a fuel droplet on the wall of a combustion chamber. In the inkjet printing and the laser-induced-forward-transfer printing, better understanding of the dynamics of droplet splashing is very useful for improving the printing quality.

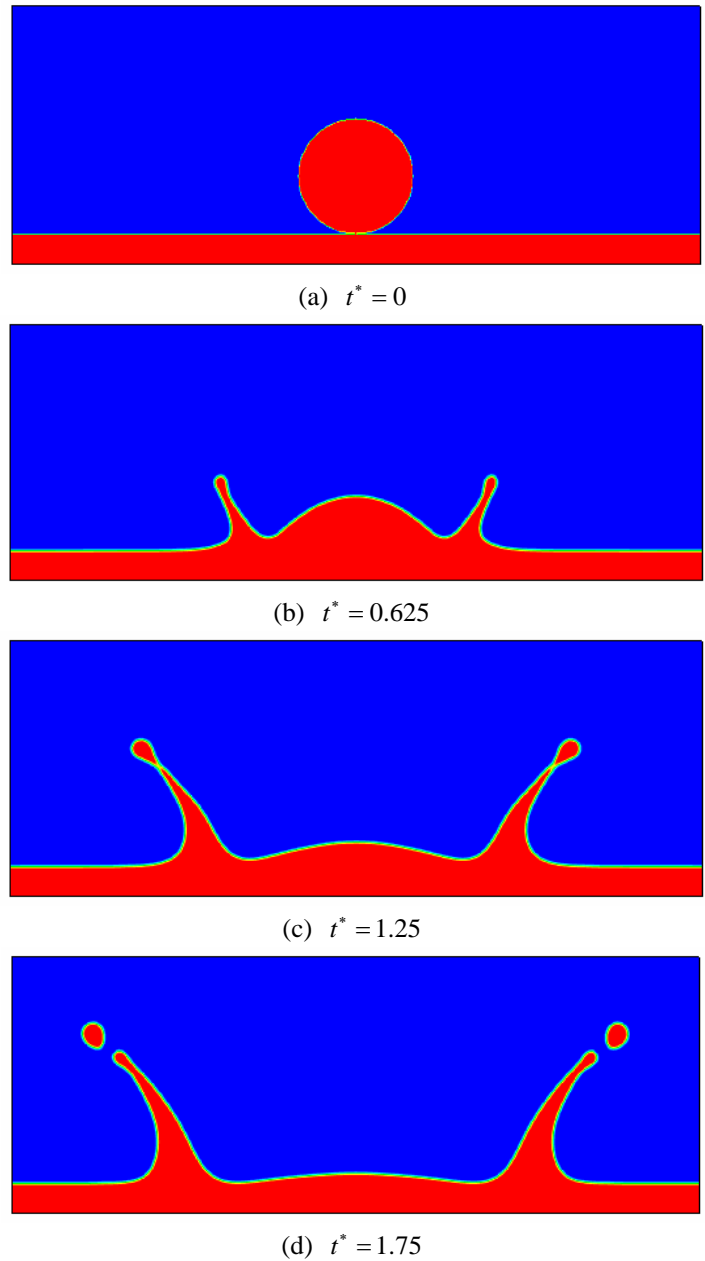

(a) $t^* = 0$

(b) $t^* = 0.625$

(c) $t^* = 1.25$

(d) $t^* = 1.75$

**Figure 5** The time evolution of droplet splashing at $1-\kappa = 1$.

In our previous work [33], we have studied the problem of droplet splashing on a thin liquid film with the Carnahan-Starling equation of state. However, the surface tension is unadjustable. In this subsection, we employ the proposed approach to investigate the effects of surface tension on droplet splashing. The computational domain is taken as $N_x \times N_y = 600 \times 225$. The liquid film is placed at the bottom of the computational domain and its height is 25 (lattice unit). The radius of the droplet is $R = 50$ and its impact velocity is $(v_x, v_y) = (0, -U)$, where $U = 0.125c$ ($c = \delta_t = 1$). The no-slip boundary condition is imposed at the bottom wall, while the open boundary condition is applied at other boundaries. The interaction potential $\psi$ is the same as that in the previous subsection and the density ratio $\rho_L/\rho_V$ is also set to be 100. The relaxation times are chosen as: $\tau_\rho = \tau_j = 1.0$, $\tau_e^{-1} = \tau_\varsigma^{-1} = 0.8$, and $\tau_q^{-1} = 1.1$. The Reynolds number is fixed at $\mathrm{Re} = UD/\upsilon_l = 1000$, in which $\upsilon_L$ is the liquid kinematic viscosity and $D$ is the diameter of the impact droplet. The viscosity ratio $\upsilon_V/\upsilon_L$ is set to be 20. Here we use $\upsilon(\rho) = \upsilon_V$ for $\rho < \rho_1$ and $\upsilon(\rho) = \upsilon_L$ for $\rho \geq \rho_1$.

First, the results of the case $\kappa = 0$ ($1 - \kappa = 1$) are shown in Fig. 5, which displays the time evolutions of the droplet and the thin liquid film from $t^* = 0$ to $t^* = 1.75$ ($t^* = Ut/D$ is the non-dimensional time). From the figure we can see that, after the impact of the droplet, a thin liquid sheet will be emitted at the intersection between the droplet and the liquid film. As time goes on, the thin liquid sheet will propagate radially away from the droplet and grow into a crown. The end rims of the crown are usually unstable and will break up into secondary droplets by the Rayleigh–Plateau instability [15], which is an important phenomenon of droplet splashing.

To show the effects of the surface tension, the density contours of the cases $1 - \kappa = 1$, 0.75, 0.25, and 0.05 at $t^* = 1.75$ are plotted in Fig. 6. The corresponding Weber number $\mathrm{We} = \rho_l U^2 D/\vartheta$ increases from 97 ($\kappa = 0$) to 1953 ($\kappa = 0.95$). From the figure we can see that, as the surface tension decreases, the emitted liquid sheet becomes thinner. Meanwhile, the tips of the liquid sheet get sharper. As a result, the end rims of the crown will break up into a number of satellite droplets: in Fig. 6(a) there is only a pair of satellite droplets, while in Fig. 6(c) two pairs of satellite droplets appear. It is also seen that the satellite droplets become smaller and smaller with the decrease of the surface tension. For the case $1-\kappa = 0.05$ whose surface tension is close to zero, some of the satellite droplets are very small, and therefore cannot be resolved with the interface thickness around five lattices. On the other hand, from Fig.

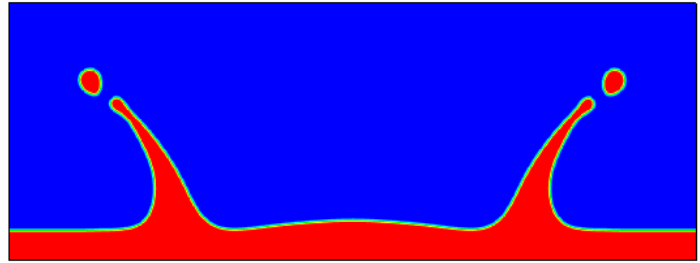

(a) $1-\kappa = 1$

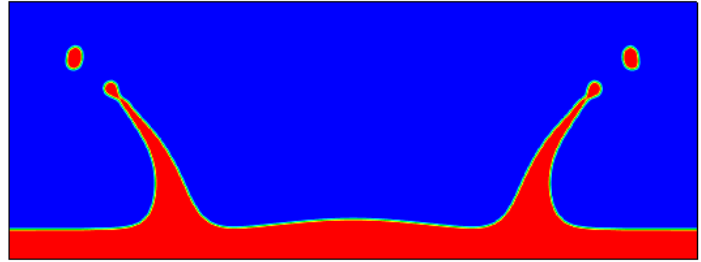

(b) $1-\kappa = 0.75$

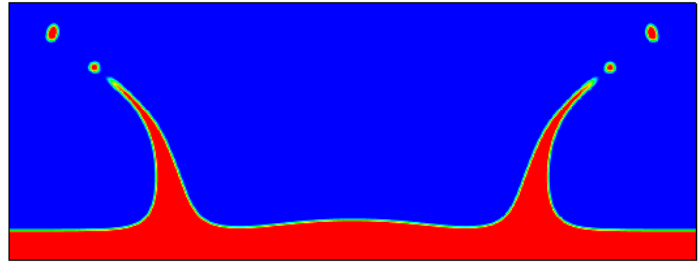

(c) $1-\kappa = 0.25$

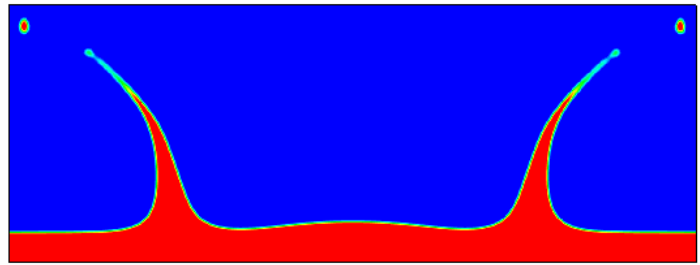

(d) $1-\kappa = 0.05$

**Figure 6** Comparison of the density contours at $t^* = 1.75$.

6(d) to Fig. 6(a) we can observe that increasing the surface tension will smooth the interface and prevent the interface pinch off. It is seen that, with the increase of the surface tension, the ends of the emitted liquid sheet get blunter due to the smoothing effect of the surface tension. According to the numerical results, it can be concluded that in droplet splashing the surface tension has important influences on the evolution of the emitted liquid sheet and the formation of secondary droplets.

To quantify the numerical results, the spread radii of the cases $1-\kappa=1$ and $1-\kappa=0.05$ at the non-dimensional time $t^*=0.05$, $0.25$, $0.75$, $1.0$, and $1.5$ are listed in Table 5, from which it can be found that there are no significant variations of the spread radius when $1-\kappa$ changes from 1.0 to 0.05, which indicates that the surface tension (and the Weber number) has a negligible effect on the spread radius. In addition, much research has shown that the spread radius obeys the power law at the early stage after the impact [40-42]. From the table we can see that for both cases the spread radius overall obeys the power law $r/D \approx 1.3\sqrt{t^*}$.

**Table 5** Comparison of the spread radius between the numerical results and the power law.

| $t^*$ | $1.3\sqrt{t^*}$ | spread radius $r/D$ | |
| --- | --- | --- | --- |
| | | $1-\kappa=1$ | $1-\kappa=0.05$ |
| 0.05 | 0.291 | 0.3100 | 0.3100 |
| 0.25 | 0.65 | 0.6728 | 0.6550 |
| 0.75 | 1.126 | 1.1708 | 1.1548 |
| 1.0 | 1.30 | 1.3317 | 1.3168 |
| 1.5 | 1.592 | 1.6034 | 1.5841 |

## VI. Conclusions

In summary, we have performed theoretical and numerical analyses for the pseudopotential LB model with the multi-range potential. It is found that the mechanical stability condition given by the multi-range potential is dependent on the parameters $G_1$ and $G_2$. As a result, the density ratio of the system will be changed when the multi-range potential is used to adjust the surface tension. In addition, it has been shown that discrete form pressure tensor of the multi-range potential contains two terms related to the surface tension. In practical applications, one term reduces the surface tension, whereas the other term enhances the surface tension, which makes the surface tension limited in a narrow range.

An alternative approach has been proposed by incorporating a source term into the LB equation and has successfully overcome the drawback of the multi-range potential. The proposed approach has the following two features. First, it guarantees that the adjustment of the surface tension does not affect the mechanical stability condition of the pseudopotential LB model, and thus allows independent adjustment of the surface tension and the density ratio. Second, as can be seen from Eq. (19), the proposed approach still retains the mesoscopic feature and the computational simplicity of the pseudopotential LB model. The new approach has been validated through numerical simulations of stationary droplets, capillary waves, and droplet splashing on a thin liquid film. The numerical results conclusively demonstrate that the proposed approach is capable of tuning the surface tension over a wide range and can keep the density ratio virtually unchanged, which may be useful for broadening the range of applicability of the pseudopotential LB model. The extensions to three-dimensional and multicomponent pseudopotential LB models will be pursued in future work.

## Acknowledgments

Support by the Engineering and Physical Sciences Research Council of the United Kingdom under Grant No. EP/I012605/1 is gratefully acknowledged.

## References

1. T. Krüger, S. Frijters, F. Günther, B. Kaoui, J. Harting, *Eur. Phys. J. Special Topics*, 2013, in press.

2. S. Unverdi and G. Tryggvason, *J. Comput. Phys.*, 1992, **100**, 25–37.

3. C.W. Hirt and B.D. Nichols, *J. Comput. Phys.*, 1981, **39**, 201–225.

4. V.E. Badalassi, H.D. Ceniceros and S. Banerjee, *J. Comput. Phys.*, 2003, **190**, 371–397.

5. J.A. Sethian and Peter Smereka, *Annu. Rev. Fluid Mech.*, 2003, **35**, 341–372.

6. S. Succi, *The Lattice Boltzmann Equation for Fluid Dynamics and Beyond*, Clarendon, Oxford, 2001.

7. M. C. Sukop and D. T. Thorne, *Lattice Boltzmann Modeling: An Introduction for Geoscientists and Engineers*, Springer, Heidelberg, Berlin, New York, 2006.

8. S. Chen and G. D. Doolen, *Annu. Rev. Fluid Mech.*, 1998, **30**, 329–364.

9. C. K. Aidun and J. R. Clausen, *Annu. Rev. Fluid Mech.*, 2010, **42**, 439–472.

10. X. Shan and H. Chen, *Phys. Rev. E*, 1993, **47**, 1815–1819; X. Shan and H. Chen, *Phys. Rev. E*, 1994, **49**, 2941–2948.

11. M. R. Swift, W. R. Osborn and J. M. Yeomans, *Phys. Rev. Lett.*, 1995, **75**, 830; M. R. Swift, E. Orlandini, W. R. Osborn and J. M. Yeomans, *Phys. Rev. E*, 1996, **54**, 5041–5052.

12. Q. Kang, D. Zhang and S. Chen, *Phys. Fluids*, 2002, **14**, 3203–3214.

13. X. He, S. Chen and R. Zhang, *J. Comput. Phys.*, 1999, **152**, 642–663.

14. X. He and G. D. Doolen, *J. Stat. Phys.*, 2002, **107**, 309–328.

15. T. Lee and C.-L. Lin, *J. Comput. Phys.*, 2005, **206**, 16–47.

16. A. Fakhari and M. H. Rahimian, *Phys. Rev. E*, 2010, **81**, 036707.

17. M. C. Sukop and D. Or, *Water Resour. Res.*, 2004, **40**, W01509.

18. Y. Gan, A. Xu, G. Zhang, and Y. Li, *Phys. Rev. E*, 2011, **83**, 056704; Y. Gan, A. Xu, G. Zhang, and Y. Li, *Commun. Theor. Phys.*, 2012, **57**, 681.

19. H. Liu, A. J. Valocchi, and Q. Kang, *Phys. Rev. E*, 2012, **85**, 046309.

20. R. Benzi, S. Chibbaro and S. Succi, *Phys. Rev. Lett.*, 2009, **102**, 026002.

21. R. Benzi, M. Sbragaglia, S. Succi, M. Bernaschi and S. Chibbaro, *J. Chem. Phys.*, 2009, **131**, 104903.

22. X. Shan, *Phys. Rev. E*, 2006, **73**, 047701.

23. M. Sbragaglia, R. Benzi, L. Biferale, S. Succi, K. Sujiyama and F. Toschi, *Phys. Rev. E*, 2007, **75**, 026702.

24. G. Falcucci, G. Bella, G. Chiatti, S. Chibbaro, M. Sbragaglia and S. Succi, *Commun. Comput. Phys.*, 2007, **2**, 1071–1084; G. Falcucci, S. Ubertinib and S. Succi, *Soft Matter*, 2010, **6**, 4357–4365.

25. M. Sbragaglia, R. Benzi, M. Bernaschi and S. Succi, *Soft Matter*, 2012, **8**, 10773–10782.

26. H. Huang, M. Krafczyk and X. Lu, *Phys. Rev. E*, 2011, **84**, 046710.

27. D. d'Humières, in *Rarefied Gas Dynamics: Theory and Simulations*, Prog. Astronaut. Aeronaut. Vol. 159, edited by B. D. Shizgal and D. P. Weaver (AIAA, Washington, D.C., 1992).

28. Z. L. Guo and C. G Zheng, *Theory and Applications of Lattice Boltzmann Method (in Chinese)*, Science Press, Beijing, 2009.

29. Q. Li, Y. L. He, G. H. Tang and W. Q. Tao, *Phys. Rev. E*, 2010, **81**, 056707.

30. X. Shan, *Phys. Rev. E*, 2008, 77, 066702.

31. P. Yuan and L. Schaefer, *Phys. Fluids*, 2006, **18**, 042101.

32. Q. Li, K. H. Luo and X. J. Li, *Phys. Rev. E*, 2012, **86**, 016709.

33. Q. Li, K. H. Luo and X. J. Li, *Phys. Rev. E*, 2013, **87**, 053301.

34. Z. Guo, C. Zheng and B. Shi, *Phys. Rev. E*, 2002, **65**, 046308.

35. M. E. McCracken and J. Abraham, *Phys. Rev. E*, 2005, **71**, 036701.

36. C. E. Colosqui, G. Falcucci, S. Ubertini and S. Succi, *Soft Matter*, 2012, **8**, 3798–3809.

37. H. Li, Y. Zhao and X. Yuan, *Soft Matter*, 2013, **9**, 1005–1009.

38. D. A. Willisa and V. Grosu, *Appl. Phys. Lett.*, 2005, **86**, 244103.

39. T. Tran, H. J. J. Staat, A. Susarrey-Arce, T. C. Foertsch, A. van Houselt, H. J. G. E. Gardeniers, A. Prosperetti, D. Lohse and C. Sun, *Soft Matter*, 2013, **9**, 3272–3282.

40. C. Josseranda and S. Zaleskib, *Phys. Fluids*, 2003, **15**, 1650–1657.

41. A. L. Yarin, *Annu. Rev. Fluid Mech.*, 2006, **38**, 159–192.

42. G. Coppola, G. Rocco, and L. de Luca, *Phys. Fluids*, 2011, **23**, 022105.

43. S. Chibbaro, G. Falcucci, G. Chiatti, H. Chen, X. Shan, S. Succi, *Phys. Rev. E*, 2008, **77**, 036705.